\documentclass[longbibliography,physrev,aps,twocolumn]{revtex4-2}
\usepackage{graphicx,epsfig}
\usepackage{amsmath,amssymb,wasysym}
\usepackage{dsfont} 
\usepackage{color}
\usepackage[linktocpage=true,colorlinks=true, pdfborder={0 0 0},linkcolor=blue,citecolor=blue,urlcolor=blue,anchorcolor=black,linkcolor=blue]{hyperref}

\usepackage{tikz} 
\usetikzlibrary{matrix}

\newcommand{\op}[1]{\hat{#1}}

\newcommand{\normspec}[1]{||{#1}||_{\text{2}}}

\newcommand{\ident}{\mathds{1}}

\newcommand{\rem}[1]{}

\newcommand{\ket}[1]{|#1\rangle}
\newcommand{\transpose}{{\text{T}}}

\newcommand{\rca}{\xi}

\newcommand{\Hs}{\op{H}}
\newcommand{\Hp}{\op{H}_{\text{p}}}

\newcommand{\order}{n}

\newcommand{\ev}{E}

\newcommand{\evEP}{\ev_{\text{EP}}}

\newcommand{\state}{\psi}

\newcommand{\stateEP}{\state_{\text{EP}}}
\newcommand{\stateEPa}{\state_{\text{A}}}
\newcommand{\stateEPb}{\state_{\text{B}}}
\newcommand{\GF}{\op{G}}

\newcommand{\ind}{l}

\newcommand{\HL}[1]{#1}

\newcommand{\projector}{\op{P}}
\newcommand{\nilpotent}{\op{N}}

\newcommand{\numsystems}{N}
\newcommand{\indsystems}{\alpha}

\newcommand{\Htl}{\op{H}'}
\newcommand{\nilpotents}{\op{n}}

\newcommand{\identa}{\mathds{1}_{\text{A}}}
\newcommand{\identb}{\mathds{1}_{\text{B}}}
\newcommand{\nilpotenta}{\op{n}_{\text{A}}}
\newcommand{\nilpotentb}{\op{n}_{\text{B}}}
\newcommand{\ordera}{\order_{\text{A}}}
\newcommand{\orderb}{\order_{\text{B}}}

\newcommand{\Ha}{\op{H}_{\text{A}}}
\newcommand{\Hb}{\op{H}_{\text{B}}}
\newcommand{\teo}{\op{K}}
\newcommand{\teoa}{\teo_{\text{A}}}
\newcommand{\teob}{\teo_{\text{B}}}
\newcommand{\rcaa}{\xi_{\text{A}}}
\newcommand{\rcab}{\xi_{\text{B}}}
\newcommand{\statea}{\state_{\text{A}}}
\newcommand{\stateb}{\state_{\text{B}}}
\newcommand{\ma}{m_{\text{A}}}
\newcommand{\mb}{m_{\text{B}}}

\newcommand{\evA}{\ev_{\text{A}}}
\newcommand{\evB}{\ev_{\text{B}}}
\newcommand{\evS}{\ev_{\Sigma}}
\newcommand{\algmul}{\alpha}
\newcommand{\algmula}{\alpha_{\text{A}}}
\newcommand{\algmulb}{\alpha_{\text{B}}}
\newcommand{\geomul}{\gamma}
\newcommand{\geomula}{\gamma_{\text{A}}}
\newcommand{\geomulb}{\gamma_{\text{B}}}

\begin{document}

\title{Higher-order exceptional points in composite non-Hermitian systems}
\author{Jan Wiersig}
\affiliation{Institut f{\"u}r Physik, Otto-von-Guericke-Universit{\"a}t Magdeburg, Postfach 4120, D-39016 Magdeburg, Germany}
\email{jan.wiersig@ovgu.de}
\author{Weijian Chen}
\affiliation{Department of Physics, North Carolina State University, Raleigh, North Carolina 27695, USA}
\date{\today}
\begin{abstract}
We show that a composite quantum system described by the tensor product of multiple systems each with a leading-order exceptional point (a non-Hermitian degeneracy at which not only eigenvalues but also eigenstates coalesce) exhibits a single leading-order exceptional point, whose order surpasses the order of any constituent exceptional point. The formation of such higher-order exceptional points does not require coupling among the subsystems. We determine explicitly the order and the spectral response strength of this exceptional point. Moreover, we observe that the energy eigenstates that do not merge are entangled. Finally, we demonstrate that general initial states disentangle during time evolution due to the presence of the higher-order exceptional point of the composite system.
\end{abstract}
\maketitle

\section{Introduction}
The time evolution of a wave system with gain and loss is described by a non-Hermitian Hamiltonian~$\op{H}$~\cite{Feshbach58,Feshbach62}, see, e.g., ultracold atoms in optical lattices~\cite{KOA97}, microwave billiards~\cite{SPK02}, one-dimensional nanostructures~\cite{CK09}, parity-time-symmetric electronics~\cite{SLL12}, compound-nucleus reactions~\cite{MRW10}, optical microcavities~\cite{Wiersig11,KYW18}, and coupled laser arrays~\cite{DM19}. Such Hamiltonians can exhibit unique singularities known as exceptional points (EPs). At an $\order$th-order EP exactly~$\order$ eigenenergies (eigenfrequencies) and the corresponding energy eigenstates (modes) coalesce~\cite{Kato66,Heiss00,Berry04,Heiss04,MA19}. This is in contrast to degeneracies of Hermitian Hamiltonians where only eigenenergies degenerate while the eigenstates can be chosen to be orthogonal. 

The physical existence of EPs has been demonstrated in a wide variety of classical settings, leading to the identification of many intriguing effects with potential applications. These include mode discrimination in multimode lasing~\cite{HMH14}, EP-enhanced sensing~\cite{Wiersig14b,COZ17,HHW17,XLK19}, mode conversion~\cite{XMJ16,DMB16}, orbital angular momentum microlasing~\cite{MZS16}, sources of circularly polarized light~\cite{RMS17}, chiral perfect absorption~\cite{SHR19}, optical amplifiers featuring an improved gain-bandwidth product~\cite{ZOE20}, and subwavelength control of light transport~\cite{XLJ23}.

While all of the above examples are classical wave systems, there has been progress of studying full quantum systems with non-Hermitian Hamiltonians by using Hamiltonian dilation~\cite{WLG19,WWY24}, postselection~\cite{NAJ19,MMC20,HWH23,HNH24}, non-Hermitian absorption spectroscopy~\cite{CLZ24}, and Bogoliubov de Gennes models~\cite{LPC18,FCV20}. Additionally, researchers investigated the non-Hermitian aspects and EPs of Liouvillian superoperators generating the time evolution of density operators~\cite{MMC19,CAH22,CLZ24}. 

Non-Hermitian systems with EPs very strongly respond to small perturbations. More concretely, a system with an EP of order $\order$ experiences an energy splitting proportional to the $\order$th root of the perturbation strength~$\varepsilon$~\cite{Kato66}. For sufficiently small~$\varepsilon$, this exceeds the linear scaling observed near a conventional degeneracy. The strength of the response to small perturbations can be characterized by a single measure: the spectral response strength~$\rca$~\cite{Wiersig22,Wiersig23b}. A system with a large $\rca$ can detect extremely small perturbations or environmental changes, making it highly sensitive and useful in particular for sensing applications~\cite{Wiersig14b,COZ17}.

The reliable fabrication of higher-order EPs is still a challenge today. Only a handful of experimental implementations are known so far, such as coupled acoustic cavities~\cite{DMX16}, photonic cavities~\cite{HHW17,LZD23}, high-dielectric spheres in the microwave regime~\cite{WHL19}, nitrogen-vacancy spin systems~\cite{WWY24}, and dissipative trapped ions~\cite{CLZ24}. Proposals for a systematic approach to achieve EPs of higher order are presented in Refs.~\cite{TEE14,KGK23}. Higher-order EPs are particularly useful for flat-top optical filters~\cite{HablerS20}, EP-enhanced sensing~\cite{HHW17,XLK19}, and in the context of entanglement generation~\cite{LCA23}. 
The latter example is a composite quantum system consisting of two non-Hermitian qubits. Mathematically, the Hilbert space of this system is a tensor product of Hilbert spaces of the single qubits. This is different to most of the higher-order EPs discussed in the literature which are constructed by a direct sum.  

The aim of this paper is to shine some light into the physics of composite non-Hermitian systems. In particular, we demonstrate that the tensor product used to combine the systems can enhance the order of EPs. We present a theory for calculating the order of the EP and its spectral response strength. The surprising effect of EP-induced disentanglement is revealed.

The structure of this paper is outlined as follows. Section~\ref{sec:composite} briefly reviews the essential basics of composite quantum systems. An illustrative example is presented in Sec.~\ref{sec:example} to demonstrate some of the peculiar properties of composite EPs. The general theory is then developed in Sec.~\ref{sec:generaltheory}.
Dynamical aspects are addressed in Sec.~\ref{sec:dynamics} and a summary is provided in Sec.~\ref{sec:summary}.

\section{Basics on composite systems}
\label{sec:composite}
The Hilbert space of a composite quantum system is the tensor product of the Hilbert spaces of the involved systems, see, e.g., Ref.~\cite{Ballentine2010}. For bipartite systems, we call them system A and B. To create operators in the full Hilbert space that act only on a single component, one has to take the tensor product of the operator acting on the subspace of interest, i.e., $\op{A}$ and $\op{B}$, with the identity operators $\identa$ and $\identb$ corresponding to the components that are to be unchanged, $\op{A}\otimes\identb$ and $\identa\otimes\op{B}$. In a matrix representation the tensor product $\otimes$ is also called the Kronecker product or matrix direct product.
If there is no interaction between systems A and B, the Hamiltonian of the composite system is given by the system Hamiltonians $\Ha$ and $\Hb$ via the Kronecker sum
\begin{equation}\label{eq:Ksum}
	\op{H} = \Ha \oplus\Hb = \Ha\otimes\identb + \identa\otimes\Hb  .
\end{equation} 
Note that the Kronecker sum is distinct from the direct sum of two matrices, despite both often being denoted by the same symbol $\oplus$. 

Equation~(\ref{eq:Ksum}) also pertains to the Liouvillian of a composite quantum system without interaction as well as to the Liouvillian of a single system if quantum jumps are ignored, see \HL{Ref.~\cite{PGS25} and} the Supplemental Material of Ref.~\cite{CAJ21}, for example.
Moreover, Hamiltonians as in Eq.~(\ref{eq:Ksum}) appear naturally in two- or higher-dimensional non-Hermitian lattices~\cite{SWW24} and for evolution matrices of higher-order operator moments~\cite{AMM21}. 

\section{An illustrative example}
\label{sec:example}
For illustration purpose, we first introduce a simplified setting given by a dimensionless toy model. Consider a system~A with a two-dimensional Hilbert space and a $2\times 2$ Hamiltonian at a second-order EP in the given basis
\begin{equation}\label{eq:exampleHa}
\Ha = \left(\begin{array}{cc}
0 & 1\\
0 & 0\\
\end{array}\right) .
\end{equation}
Such a Hamiltonian can appear in optical microdisks and microrings with fully asymmetric backscattering~\cite{Wiersig18b}. The eigenvalue is $0$ with algebraic multiplicity of $2$ and the only eigenstate is $\ket{\statea} = (1,0)^\transpose$ with geometric multiplicity of $1$; the superscript~$\transpose$ indicates the transpose. Now, we consider a system~B with, for simplicity, exactly the same Hamiltonian $\Hb = \Ha$ at an EP of order $2$ having eigenvalue $0$ and eigenstate~$\ket{\stateb} = (1,0)^\transpose$. 

The composite system with Hamiltonian in Eq.~(\ref{eq:Ksum}) is (after a unitary transformation) equivalent to the one in Ref.~\cite{LCA23} but computationally more convenient. Does the composite system contain an EP? And if so, what is its order? The answer given in Ref.~\cite{LCA23} is 4 and there is a simple and seemingly safe argument for it (which, however, has not been used in Ref.~\cite{LCA23}): each of the Hamiltonians $\Ha$ and $\Hb$ possess exactly one eigenstate (the geometric multiplicity is 1), $\ket{\statea}$ and $\ket{\stateb}$, and therefore the only tensor product to form is
\begin{equation}\label{eq:exampleEPv}
\ket{\stateEP} = \ket{\statea}\otimes\ket{\stateb} = \left(\begin{array}{c}
1\\
0\\
0\\
0\\
\end{array}\right) .
\end{equation}
As a result, the geometric multiplicity is 1. Together with an algebraic multiplicity of 4 (all eigenvalues are zero) this indicates an EP of order 4. However, direct calculation of the Hamiltonian in Eq.~(\ref{eq:Ksum}) proves that this hasty conclusion is incorrect:
\begin{equation}\label{eq:exampleH}
\op{H} = \left(\begin{array}{cccc}
0 & 1 & 1 & 0\\
0 & 0 & 0 & 1\\
0 & 0 & 0 & 1\\
0 & 0 & 0 & 0\\
\end{array}\right)  .
\end{equation}
The eigenvalues are $0$ with algebraic multiplicity of 4 and there are \emph{two} eigenstates
\begin{equation}\label{eq:exampleEPv0}
\ket{\stateEP} = \left(\begin{array}{c}
1\\
0\\
0\\
0\\
\end{array}\right)\;,
\ket{\state_0} = \left(\begin{array}{c}
	0\\
	-1\\
	1\\
	0\\
\end{array}\right)
\end{equation}
up to a phase factor and normalization. The additional eigenstate $\ket{\state_0}$ is not a tensor product of the EP eigenstates of systems A and B; in fact, it is a (maximally) entangled state. The geometric multiplicity is therefore equal to~2. To investigate this further, we employ the unitary matrix
\begin{equation}\label{eq:exampleU}
\op{U} := \left(\begin{array}{cccc}
0 & \frac{1}{\sqrt{2}} & -\frac{1}{\sqrt{2}} & 0\\
1 & 0                  & 0                   & 0\\
0 & \frac{1}{\sqrt{2}} & \frac{1}{\sqrt{2}}  & 0\\
0 & 0                  & 0                   & 1\\
\end{array}\right) 
\end{equation}
to transform the Hamiltonian $\op{H}$ in Eq.~(\ref{eq:exampleH}) to a Jordan normal form
\begin{equation}\label{eq:exampleJNF}
\op{U}\op{H}\op{U}^\dagger = \sqrt{2}\begin{tikzpicture}[baseline]
\matrix (M1) [matrix of nodes,{left delimiter=(},{right delimiter=)}] 
{
0 & 0 & 0 & 0\\
0 & 0 & 1 & 0\\
0 & 0 & 0 & 1\\
0 & 0 & 0 & 0\\
};
\draw[red,dashed,thick]
(M1-2-2.north west) -| (M1-4-4.south east) -| (M1-2-2.north west);
\end{tikzpicture}.
\end{equation}
This clearly demonstrates that the composite system exhibits a more complex degeneracy (see for such complex degeneracies also Refs.~\cite{WWH20,Znojil20,GOH21,KGK23,BS25,NCM25}) consisting of a third-order EP and a separate eigenstate, each with a zero eigenvalue.
The result is consistent with Ref.~\cite{AMM21} which does not consider composite quantum systems but evolution matrices of higher-order operator moments of the form as in Eq.~(\ref{eq:Ksum}) with $\Ha = \Hb$ at a second-order EP.

It is worth taking a moment to think through the original argument of Ref.~\cite{LCA23} for the order of the EP in the composite system. To do so, consider the perturbed Hamiltonians
\begin{equation}\label{eq:exampleHap}
\Ha = \left(\begin{array}{cc}
0 & 1\\
\varepsilon & 0\\
\end{array}\right) = \Hb .
\end{equation}
These Hamiltonians are not at an EP for $\varepsilon \neq 0$. The eigenvalues are $E_\pm = \pm\sqrt{\varepsilon}$ and the eigenstates are $\ket{\state_\pm} = (1,\pm\sqrt{\varepsilon})^\transpose$. The tensor product of these states gives the eigenstates (up to normalization) of the Hamiltonian in Eq.~(\ref{eq:Ksum}) with zero eigenvalue
\begin{equation}\label{eq:example_v01}
	\ket{\state_0} = \left(\begin{array}{c}
		1\\
		-\sqrt{\varepsilon}\\
		\sqrt{\varepsilon}\\
		-\varepsilon\\
	\end{array}\right)\,,\;
	\ket{\state_1} = \left(\begin{array}{c}
		1\\
		\sqrt{\varepsilon}\\
		-\sqrt{\varepsilon}\\
		-\varepsilon\\
	\end{array}\right)
\end{equation}
and the eigenstates with eigenvalues $2\sqrt{\varepsilon}$ and $-2\sqrt{\varepsilon}$
\begin{equation}\label{eq:example_v023}
	\ket{\state_2} = \left(\begin{array}{c}
		1\\
		\sqrt{\varepsilon}\\
		\sqrt{\varepsilon}\\
		\varepsilon\\
	\end{array}\right)\,,\;
	\ket{\state_3} = \left(\begin{array}{c}
		1\\
		-\sqrt{\varepsilon}\\
		-\sqrt{\varepsilon}\\
		\varepsilon\\
	\end{array}\right)  .
\end{equation}
In the limiting case $\varepsilon\to 0$ all four eigenstates converge to the EP eigenstate in Eq.~(\ref{eq:exampleEPv}). This appears to be an unambiguous signature of a fourth-order EP. The fallacy lies in the eigenvalue degeneracy of the eigenstates in Eq.~(\ref{eq:example_v01}), which allows for alternative linear combinations such as
\begin{equation}
	\ket{\tilde{\state}_0} = \left(\begin{array}{c}
		0\\
		-1\\
		1\\
		0\\
	\end{array}\right)\,,\;
	\ket{\tilde{\state}_1} = \left(\begin{array}{c}
	1\\
	0\\
	0\\
	-\varepsilon\\
\end{array}\right)  .
\end{equation}
The eigenstate $\ket{\tilde{\state}_0}$ does not converge to the EP eigenstate in Eq.~(\ref{eq:exampleEPv}). Instead, it is equal to the isolated eigenstate in Eq.~(\ref{eq:exampleEPv0}). This observation is quite remarkable. It demonstrates that the more complex degeneracies consisting of an EP and other states can complicate analysis and the physical interpretation. Our general theory, which will be presented in the next section, is immune to the pitfalls presented above.

\section{General theory}
\label{sec:generaltheory}
\subsection{Preliminaries}
A non-Hermitian $m\times m$ Hamiltonian can be expanded as~\cite{Kato66}
\begin{equation}\label{eq:Hexpand}
	\op{H} = \sum_\ind (E_\ind\projector_\ind+ \nilpotent_\ind)
\end{equation}
where~$\ind = 1,2,\ldots$ runs over the relevant part of the point spectrum $E_\ind$ including isolated energy eigenvalues and EPs. The operators~$\projector_\ind$ are projectors onto the generalized eigenspaces of the corresponding eigenvalues~$\ev_\ind$ with
\begin{equation}\label{eq:pp}
	\projector_j\projector_\ind = \delta_{j\ind}\projector_\ind 
\end{equation}
and Kronecker delta $\delta_{j\ind}$. The $\projector_\ind$ are not orthogonal projectors in general, i.e., $\projector_\ind \neq \projector^\dagger_\ind$.   
The operators~$\nilpotent_\ind$ for a given EP of order $\order_\ind \geq 2$ are nilpotent operators of index~$\order_\ind$; hence, $\nilpotent_\ind^{\order_\ind} = 0$ but $\nilpotent_\ind^{\order_\ind-1} \neq 0$. It holds that
\begin{equation}\label{eq:pn}
	\projector_\ind \nilpotent_\ind = \nilpotent_\ind\projector_\ind = \nilpotent_\ind  .
\end{equation}
The $m\times m$ Green's function of the general Hamiltonian~$\Hs$ can be expanded as~\cite{Kato66}
\begin{equation}\label{eq:GKato}
	\GF(\ev) = \sum_\ind\left[\frac{\projector_\ind}{\ev-\ev_\ind} + \sum_{k=2}^{\order_\ind} \frac{\nilpotent_\ind^{k-1}}{(\ev-\ev_\ind)^k}\right]  .
\end{equation}
For an energy $\ev\approx\evEP$ close to an EP energy $\evEP = \ev_{\ind}$, the contribution of the Green's function with $k=\order_{\ind}$ is the dominant one. 
Using this  Green's function it has been shown in Ref.~\cite{Wiersig23b} that 
for a generic perturbation $\op{H}+\varepsilon\Hp$ with small perturbation strength $\varepsilon > 0$, the following inequality holds for the split eigenvalues $\tilde{\ev}_j$, $j = 1,\ldots,\order_\ind$, of the perturbed Hamiltonian
\begin{equation}\label{eq:ub}
	|\tilde{\ev}_j-\evEP|^{\order_\ind} \leq \varepsilon \normspec{\Hp}\,\rca 
\end{equation}
with the spectral norm (operator norm) $\normspec{\cdot}$~\cite{Johnston21} and the so-called spectral response strength 
\begin{equation}\label{eq:rcagen}
	\rca = \normspec{\nilpotent_\ind^{\order_\ind-1}}  .
\end{equation}
The spectral response strength is a measure of how strongly a non-Hermitian system with an EP can react to a generic perturbation~\cite{Wiersig22}. It has been calculated and discussed for various systems~\cite{Wiersig22b,Wiersig22c}. The relation to the Petermann factor is discussed in Refs.~\cite{Wiersig23,KWS25}. Recently, a scheme for the computation of the spectral response strength directly from numerical wave simulations has been presented~\cite{KW25}.

\subsection{Bipartite systems}
\label{sec:bipartite}
\subsubsection{Existence of an higher-order EP}
For the sake of simplicity, we initially consider a composite quantum system consisting of two quantum systems A and B. The Hilbert space of the former is $\ma$-dimensional and that of the latter is $\mb$-dimensional. The tensor product of these two spaces gives the Hilbert space of the composite system with dimension $m = \ma\mb$. The $m\times m$ Hamiltonian of the composite system is given by the Kronecker sum in Eq.~(\ref{eq:Ksum}) with $\Ha$ being the $\ma\times\ma$-Hamiltonian and $\identa$ the $\ma\times\ma$-identity of the system A and analog for $\Hb$ and $\identb$. 

One of our basic assumptions is that the relevant part of the spectrum of systems A and B is completely degenerated. \HL{The eigenvalue of each subsystem, $\evA$ and $\evB$, can be different}. With the above assumption, the algebraic multiplicity of the degeneracy is $\algmula = \ma$ for system~A and $\algmulb = \mb$ for system B. Using the fact that the eigenvalues of a matrix $\op{A} = \op{B}\oplus\op{C}$ are given by $b_i+c_j$ where $b_i$ are the eigenvalues of $\op{B}$ and $c_j$ are the eigenvalues of~$\op{C}$~\cite{Johnston21} we conclude that the only eigenvalue of~$\op{H}$ is \HL{$\evS = \evA+\evB$} with algebraic multiplicity $\algmul = m$.

Our second basic assumption is that each degeneracy exhibits a dominant EP. The order of the EP, $2 \leq \ordera \leq \ma$ and $2\leq \orderb\leq \mb$, can be different. We place no limitations on the geometric multiplicity of each degeneracy, other than the apparent constraints $\geomula \leq \algmula - \ordera + 1$ and $\geomulb \leq \algmulb - \orderb + 1$, which result from the possible existence of additional (lower-order) EPs. 

We introduce the traceless part of the Hamiltonians 
\HL{
\begin{eqnarray} 
\nilpotenta & := & \Ha-\evA\identa ,\\	
\nilpotentb & := & \Hb-\evB\identb  ,\\
\label{eq:Htl}
\Htl        & := & \op{H}-\evS\ident
\end{eqnarray} 
} with the identity $\ident=\identa\otimes\identb$ for the composite system. 
As the spectrum of system A is degenerate with a dominant EP of order $\ordera$, the matrix~$\nilpotenta$ is nilpotent of order~$\ordera$. The same logic implies that $\nilpotentb$ is nilpotent of order $\orderb$. 

In terms of the traceless parts, Eq.~(\ref{eq:Ksum}) can be rewritten as
\begin{equation} 
	\Htl = \nilpotenta\otimes\identb + \identa\otimes\nilpotentb . 	
\end{equation} 
With the basic rules of the tensor product~\cite{Johnston21}, we calculate monomials, such as
\begin{equation} 
	(\Htl)^2 = \nilpotenta^2\otimes\identb + 2\nilpotenta\otimes\nilpotentb +\identa\otimes\nilpotentb^2  . 	
\end{equation} 
Using the binomial coefficient and the nilpotency of $\nilpotenta$ and $\nilpotentb$, it follows with the auxiliary quantity $\order := \ordera+\orderb-1$ that
\begin{equation}\label{eq:Htlnm1}
(\Htl)^{\order-1} = \frac{(\order-1)!}{(\ordera-1)!(\orderb-1)!}\nilpotenta^{\ordera-1}\otimes\nilpotentb^{\orderb-1} 
\end{equation} 
and 
\begin{equation}\label{eq:Htln}
(\Htl)^{\order} = 0  . 	
\end{equation} 
Hence, $\Htl$ is nilpotent of order $\order$. Note that the binomial expansion~(\ref{eq:Htlnm1}) contains only one term, as all other terms vanish because either $\nilpotenta$ or $\nilpotentb$ appear with an exponent greater than $\ordera-1$ or $\orderb-1$, respectively.

Given that the eigenvalues of $\Htl$ are zero, the expansion in Eq.~(\ref{eq:Hexpand}) is
\begin{equation}\label{eq:Htlexpand}
	\Htl = \sum_\ind \nilpotent_\ind .
\end{equation}
From Eqs.~(\ref{eq:pp}) and (\ref{eq:pn}) follow
\begin{equation}\label{eq:nnindep}
\nilpotent_j\nilpotent_\ind = 0
\;\; \text{if}\; j\neq\ind 
\end{equation}
which we then apply in Eq.~(\ref{eq:Htlexpand}) to obtain
\begin{equation}\label{eq:Htlexpandnm1}
	(\Htl)^{\order-1} = \sum_\ind \nilpotent_\ind^{\order-1} .
\end{equation}
Using the fact that $\nilpotenta^{\ordera-1}$ and $\nilpotentb^{\orderb-1}$ are rank-1 matrices and that the tensor product of two rank-1 matrices is a rank-1 matrix~\cite{Johnston21}, we conclude that the right-hand side (RHS) of Eq.~(\ref{eq:Htlnm1}) is a rank-1 matrix. And so is the left-hand side. This in turn implies that the RHS of Eq.~(\ref{eq:Htlexpandnm1}) must be of rank 1 as well. Moreover, since each of the nonzero $\nilpotent_\ind^{\order-1}$ is of rank 1 and the $\nilpotent_\ind$ are linearly independent because of Eq.~(\ref{eq:nnindep}), only one such term on the RHS of Eq.~(\ref{eq:Htlexpandnm1}) exists. Hence, in the expansion~(\ref{eq:Hexpand}) is an operator $\nilpotent_\ind$ being nilpotent of order $\order_\ind = \order$. This leads to one of our key results: the composite system has a dominant EP of order 
\begin{equation}\label{eq:orderbipartite}
\order = \ordera+\orderb-1
\end{equation}
with eigenvalue \HL{$\evA+\evB$}. Since $\ordera$ and $\orderb$ are larger than or equal to~2, the order $\order$ is larger than both $\ordera$ and $\orderb$.
For determining the eigenstate of this EP, we mention that the image of operators such as~$\nilpotenta^{\ordera-1}$ and $\nilpotentb^{\orderb-1}$ is the respective EP eigenstate, i.e., $\ket{\stateEPa}$ and $\ket{\stateEPb}$; see, e.g., Ref.~\cite{Wiersig22}. From Eq.~(\ref{eq:Htlnm1}) we then conclude
\begin{equation}
\ket{\stateEP} = \ket{\stateEPa}\otimes\ket{\stateEPb}  .
\end{equation}

Additional EPs may be present, but their order must be smaller than~$\order$. The geometric multiplicity therefore obeys 
\begin{equation}\label{eq:gleq}
\geomul \leq \algmul - \order +1 .
\end{equation}
The number of eigenstates of $\op{H}$ that can be written as a tensor product of the Hilbert spaces of systems A and B is $\geomula\geomulb$. If $\geomul >\geomula\geomulb$ there are additional eigenstates that must be entangled.

In our illustrative example in Sec.~\ref{sec:example} we have for systems A and B the dimensions $\ma = \mb = 2$, the EP orders $\ordera = \orderb = 2$, and the multiplicities $\algmula = \algmulb = 2$ and $\geomula = \geomulb = 1$. According to Eq.~(\ref{eq:orderbipartite}) the resulting higher-order EP has the order $\order = 3$. Since $\order$ is smaller than the algebraic multiplicity $\algmul = \algmula\algmulb = 4$, the geometric multiplicity $\geomul$ must be larger than 1 and with inequality~(\ref{eq:gleq}) we conclude $\geomul = 2 > \geomula\geomulb = 1$. Hence, there is one tensor-product EP eigenstate and one entangled eigenstate. This is entirely in agreement with the direct calculation in Sec.~\ref{sec:example}.

It is noted that the class of systems considered here differs from that in Ref.~\cite{TEE14}. In the latter, two \emph{coupled} bosonic oscillators with gain and loss are tuned to a second-order EP. Populating them with photons leads to higher-order EPs.

\subsubsection{Spectral response strength}
In the previous subsection, we determined 
\begin{equation}
\nilpotent_\ind^{\order-1} = \frac{(\order-1)!}{(\ordera-1)!(\orderb-1)!}\nilpotenta^{\ordera-1}\otimes\nilpotentb^{\orderb-1}  .
\end{equation} 
Using the fact that the spectral norm of a Kronecker product is the product of the spectral norms of each factor~\cite{Johnston21} we write
\begin{equation}
	\normspec{\nilpotent_\ind^{\order-1}} = \frac{(\order-1)!}{(\ordera-1)!(\orderb-1)!}\normspec{\nilpotenta^{\ordera-1}}\normspec{\nilpotentb^{\orderb-1}}  .
\end{equation} 
With the spectral response strengths [Eq.~(\ref{eq:rcagen})] of systems~A and B, $\rcaa = \normspec{\nilpotenta^{\ordera-1}}$ and $\rcab = \normspec{\nilpotentb^{\orderb-1}}$, we find for the spectral response strength of the EP in the composite system
\begin{equation}\label{eq:rcabipartite}
\rca = (\order-1)!\frac{\rcaa}{(\ordera-1)!}\frac{\rcab}{(\orderb-1)!}  .
\end{equation} 
This is the second key result of this paper. It allows a straightforward computation of the spectral response strength of the leading-order EP in the composite system once the spectral response strengths of the leading-order EPs in the individual systems are known.

\HL{Keep in mind that the spectral response strength in Eq.~(\ref{eq:rcabipartite}) quantifies the response of the composite system to a generic perturbation. In a bipartite system that we discuss here, a necessary condition for a perturbation to be generic is that it involves both systems, i.e., it describes an interaction. Without such an interaction the higher-order EP has no influence on the eigenvalue splitting and the spectral response is described by $\rcaa$ and~$\rcab$ alone.}

\subsection{Multipartite systems}
\label{sec:multipartite}
The results of the previous subsections can be straightforwardly generalized to systems consisting of $\numsystems\geq 2$ systems. Again our basic assumptions are that each system labeled here by $\indsystems$ is completely degenerate \HL{with eigenvalue $\ev_\indsystems$ and exhibits a dominant EP of order $\order_\indsystems$; $\indsystems = 1,\ldots,\numsystems$.} We denote the traceless part of the Hamiltonian $\op{H}_\indsystems$ by~$\nilpotents_\indsystems$. In accordance with the reasoning outlined in Sec.~\ref{sec:bipartite} we find that the composite system has an EP with eigenvalue \HL{$\evS = \sum_\indsystems\ev_\indsystems$} and with the multinominal coefficient we get
\begin{equation}\label{eq:Ngeneral}
	\nilpotent_\ind^{\order-1} = \frac{(\order-1)!}{\prod_{\indsystems=1}^\numsystems (\order_\indsystems-1)!}\;\bigotimes_{\indsystems=1}^\numsystems\nilpotents_\indsystems^{\order_\indsystems-1} 
\end{equation} 
where $\order := 1 + \sum_\indsystems (\order_\indsystems-1)$. Hence, the composite system exhibits a dominant EP of order 
\begin{equation}\label{eq:ordermultipartite}
\order = 1 + \sum_{\indsystems=1}^\numsystems (\order_\indsystems-1)  . 
\end{equation}
This result is consistent with Refs.~\cite{AMM21,SSZ25} which treated the special case where all EPs are of second order. 

For the EP eigenstate we find
\begin{equation}\label{eq:EPtp}
\ket{\stateEP} = \bigotimes_{\indsystems=1}^\numsystems\ket{\state_{\indsystems}} 
\end{equation}
where $\ket{\state_{\indsystems}}$ is the leading-order EP-eigenstate of $\op{H}_\indsystems$.
 
Again, we exploit that the spectral norm of a Kronecker product is the product of the spectral norms of each factor~\cite{Johnston21} to obtain from Eq.~(\ref{eq:Ngeneral}) the spectral response strength of the EP [Eq.~(\ref{eq:rcagen})] in the composite system
\begin{equation}\label{eq:rcageneral}
\rca = (\order-1)! \;\prod_{\indsystems=1}^\numsystems\frac{\rca_\indsystems}{(\order_\indsystems-1)!}  .
\end{equation} 
Here $\rca_\indsystems = \normspec{\nilpotents_\indsystems^{\order_\indsystems-1}}$ is the spectral response strength of the respective individual leading-order EP.

\HL{
It is worth mentioning that all the results for multipartite systems presented in this section can also be directly inferred from the results for bipartite systems derived in Sec.~\ref{sec:bipartite} through the method of mathematical induction. As an example, we briefly explain how Eq.~(\ref{eq:ordermultipartite}) can be derived from Eq.~(\ref{eq:orderbipartite}). We begin by examining the base case of $N=2$  where both equations coincide, thereby validating Eq.~(\ref{eq:ordermultipartite}) for this simple case. Next, we demonstrate the inductive step: assuming Eq.~(\ref{eq:ordermultipartite}) holds for $N=k$, we show it also holds for $N=k+1$. To achieve this, we group all systems with $\alpha \leq k$  into a composite system and determine its EP order using Eq.~(\ref{eq:ordermultipartite}). This composite system, combined with the remaining system, forms a bipartite system, which we then analyze using  Eq.~(\ref{eq:orderbipartite}). The resulting EP order agrees with  Eq.~(\ref{eq:ordermultipartite}) for $N=k+1$. This completes the induction argument.
}

\section{Dynamics}
\label{sec:dynamics}
In this section we demonstrate that the dynamics in a composite system made of degenerate systems with EPs gives rise to an interesting and counterintuitive effect.
As discussed in the previous sections, the Hamiltonian $\Hs$ of the composite system exhibits a dominant EP with order~$\order$, eigenstate $\ket{\stateEP}$, and eigenvalue \HL{$\evEP = \evS$}. 

We consider in the following the non-unitary time-evolution operator 
\begin{equation}\label{eq:teo}
\teo(t) := e^{-\frac{i}{\hbar}\Hs t}
\end{equation} 
which maps the initial state $\ket{\psi(0)}$ to the time-evolved state $\ket{\psi(t)} = \teo(t)\ket{\psi(0)}$. 
We again define the traceless part of the Hamiltonian $\Htl$ as in Eq.~(\ref{eq:Htl}). In Sec.~\ref{sec:bipartite} it was shown that $\Htl$ is nilpotent of order~$\order$. This implies that the expansion of the following exponential function is truncated to its first $\order$ terms
\begin{equation}
	e^{-\frac{i}{\hbar}\Htl t} = \sum_{j=0}^{\order-1}\frac{1}{j!}\left(-\frac{i}{\hbar}\Htl t\right)^{j} .
\end{equation}
Next, using the fact that all eigenvalues of $\Hs$ are equal to $\evEP$ we can write the time-evolution operator as
\begin{equation}
	\teo(t) = e^{-i\frac{\evEP}{\hbar} t}\sum_{j=0}^{\order-1}\frac{(-i t)^{j}}{j!\hbar^j} \Htl^{j}  .
\end{equation}
For a generic initial state, the relevant term for long times~$t$ is
\begin{equation}\label{eq:Uasym}
	\teo(t) = e^{-i\frac{\evEP}{\hbar} t}\frac{(-i t)^{\order-1}}{(\order-1)!\hbar^{\order-1}} \Htl^{\order-1}  .
\end{equation}
This result is in line with the studies of dynamics at EPs in Refs.~\cite{Longhi18b,Wiersig22}.

Next, we use again the fact that the image of $\Htl^{\order-1}$ is the eigenstate $\ket{\stateEP}$, which is a tensor product of the EP states of the subsystems; see Eq.~(\ref{eq:EPtp}). Hence for long times the image of $\teo(t)$ approaches $\ket{\stateEP}$. This let us conclude that the state of the composite system $\ket{\state}$ approaches the state of the dominant EP. Remarkable is that the dynamics in general starts with an entangled state but ends in a non-entangled state. We call this effect \emph{EP-induced disentanglement}. It is important to emphasize that this effect is not related to the exceptional phase transition at an EP observed in Ref.~\cite{HWH23} and is not in contradiction with Ref.~\cite{LCA23} reporting a speeding up of entanglement \emph{near} an EP.

\begin{figure}[ht]
	\includegraphics[width=0.95\columnwidth]{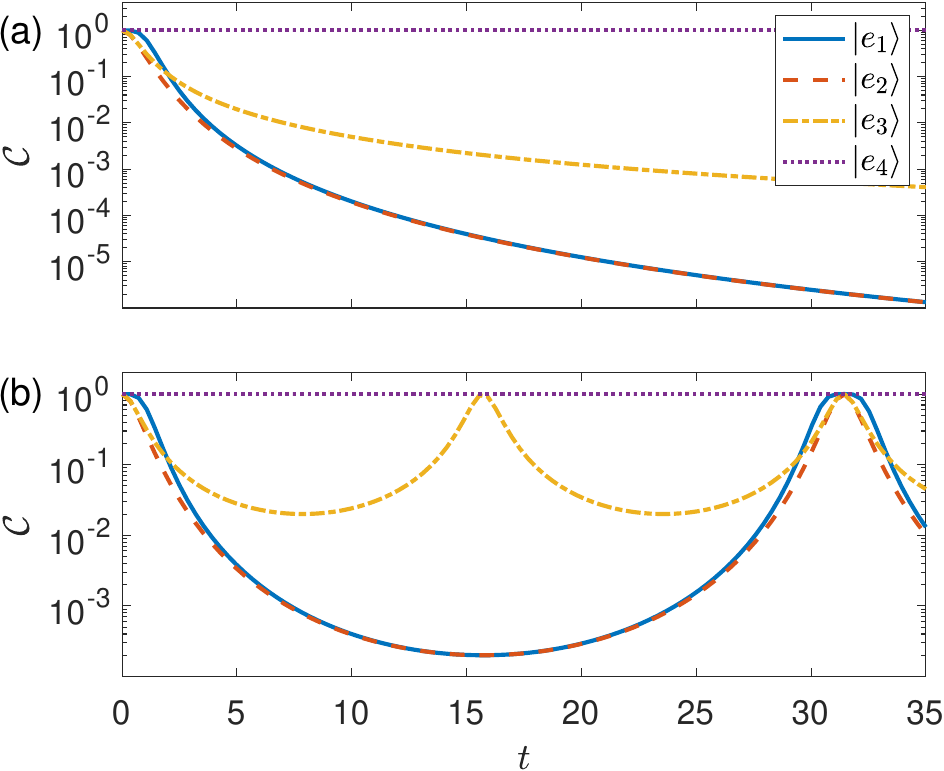}
	\caption{Time evolution of concurrence~${\cal{C}}$ [Eq.~(\ref{eq:C})] for four initial Bell states~$\ket{e_j}$ [Eqs.~(\ref{eq:e12}) and (\ref{eq:e34})] in the composite non-Hermitian system defined by Eqs.~(\ref{eq:Ksum}) and~(\ref{eq:exampleHap}). Time $t$ is in dimensionless units. (a) $\varepsilon = 0$, i.e., the system is at the EP. (b) $\varepsilon = 0.01$, i.e., the system is not at the EP but close to it.}
	\label{fig:eps0}
\end{figure}

To better illustrate this, we refer to our example in Sec.~\ref{sec:example}. To be more specific, we consider Eqs.~(\ref{eq:Ksum}) and~(\ref{eq:exampleHap}). Following Ref.~\cite{LCA23} we employ the concurrence ${\cal{C}}$ as a convenient measure of entanglement. For a normalized state $\ket{\psi} = (\alpha,\zeta,\beta,\delta)^\transpose$ the concurrence is defined by
\begin{equation}\label{eq:C}
{\cal{C}} := 2|\alpha\delta-\beta\zeta|  .
\end{equation}
As initial conditions we choose the four Bell states (see Ref.~\cite{LCA23})
\begin{equation}\label{eq:e12}
\ket{e_1} = \frac{1}{\sqrt{2}}\left(\begin{array}{c}
	1\\
	0\\
	0\\
	1\\
\end{array}\right)\,,\;
	\ket{e_2} = \frac{i}{\sqrt{2}}\left(\begin{array}{c}
		1\\
		0\\
		0\\
		-1\\
	\end{array}\right)
\end{equation}
and
\begin{equation}\label{eq:e34}
	\ket{e_3} = \frac{i}{\sqrt{2}}\left(\begin{array}{c}
		0\\
		1\\
		1\\
		0\\
	\end{array}\right)\,,\;
	\ket{e_4} = \frac{1}{\sqrt{2}}\left(\begin{array}{c}
		0\\
		-1\\
		1\\
		0\\
	\end{array}\right)  .
\end{equation}
For these maximally entangled states the concurrence ${\cal C}$ is unity.
Figure~\ref{fig:eps0}(a) shows that at the EP ${\cal{C}}$ decays for all Bell states as predicted by the phenomenon of EP-induced disentanglement, except for $\ket{e_4}$ where~${\cal{C}}$ stays constant. The latter is consistent with the discussion around Eq.~(\ref{eq:exampleEPv0}) in Sec.~\ref{sec:example} as $\ket{e_4}$ is an eigenstate of the Hamiltonian. 
	
Slightly away from the EP, the situation can change drastically (here for $\varepsilon > 0$), as shown in Fig.~\ref{fig:eps0}(b). The concurrence recovers, reaching values up to unity. \HL{A similar phenomenon is studied in Ref.~\cite{LCA23} with a weak subsystem coupling, which is referred to as the speeding up of entanglement in the proximity of an EP. To see that this is not in contradiction to our finding, consider the Bell state~$\ket{e_3}$ defined in Eq.~(\ref{eq:e34}). It can be expanded into the energy eigenstates $\ket{\psi_2}$ and $\ket{\psi_3}$ with eigenvalues $2\sqrt{\varepsilon}$ and $-2\sqrt{\varepsilon}$, cf. Eq.~(\ref{eq:example_v023}), as $\ket{e_3} \propto \ket{\psi_2} - \ket{\psi_3}$. When this state evolves in time, the superposition leads to concurrence oscillations with a period proportional to $1/\sqrt{\epsilon}$ for $\varepsilon > 0$. For sufficiently large $\varepsilon$ this gives a speeding up of entanglement. We consider here the opposite limit of $\varepsilon \to 0^+$ where the time scale for concurrence returning to unity goes to infinity, which effectively leads to disentanglement. This is the EP-induced disentanglement discussed in the present paper.}

For $\varepsilon < 0$ we do not observe the recovering of the concurrence (not shown) which again is consistent with Ref.~\cite{LCA23}. \HL{The system now is in a parity-time-symmetry broken regime (see, e.g., Ref.~\cite{FE17}). The real part of all four eigenvalues is zero, which implies that there are no oscillations. The nonzero imaginary part quantifies gain and losses of the eigenstates. Taking the same example of the Bell state~$\ket{e_3}$, it evolves to the energy eigenstate~$\ket{\psi_2}$ with concurrence ${\cal C} = 0$ due to this gain-loss imbalance. This is a fundamentally different mechanism compared to EP-induced disentanglement.}

Finally, let us mention that Eq.~(\ref{eq:Uasym}) can be used for an alternative proof of Eq.~(\ref{eq:orderbipartite}). To do so, we utilize the property of the Kronecker sum~\cite{Neudecker69}
\begin{equation}
	e^{\op{A}\oplus\op{B}} = e^{\op{A}}\otimes e^{\op{B}}  .
\end{equation}
From Eq.~(\ref{eq:Ksum}) then follows for the time evolution operator~(\ref{eq:teo}) of the composite system
\begin{equation}\label{eq:UUaUb}
	\teo(t) = e^{-\frac{i}{\hbar}\Ha t} \otimes e^{-\frac{i}{\hbar}\Hb t} = \teoa(t)\otimes\teob(t)
\end{equation}
where $\teoa(t)$ and $\teob(t)$ are the time-evolution operators of system A and B. From Eq.~(\ref{eq:Uasym}) we conclude that~$\teoa(t)$ behaves asymptotically as $t^{\ordera-1}$ and $\teob(t)$ as~$t^{\orderb-1}$. Hence, according to Eq.~(\ref{eq:UUaUb}), $\teo(t)$ behaves asymptotically as~$t^{\ordera+\orderb-2}$. This is in agreement with an EP of order $\order$ given by Eq.~(\ref{eq:orderbipartite}), see also Ref.~\cite{Wiersig20}.

\section{Summary}
\label{sec:summary}
Our study on composite non-Hermitian systems formed by a tensor product of degenerate systems has revealed a number of interesting findings to be summarized in this section.

If each individual system exhibits a leading-order exceptional point of some order $\order_\indsystems$ then the composite system exhibits a single leading-order exceptional point with order~$\order$ [Eq.~(\ref{eq:ordermultipartite})] larger than any of the $\order_\indsystems$. This is an interesting mechanism for creating higher-order exceptional points that might have applications, e.g., for sensing. A perturbation that involves all subsystems---a global perturbation---can be a generic perturbation yielding a large splitting with~$\varepsilon^{1/\order}$ scaling. In contrast, local perturbations restricted to \HL{two subsystems or just a single subsystem lead to a smaller splitting, for example, in the case of a single subsystem with an} $\varepsilon^{1/\order_\indsystems}$ scaling. Hence a sensor device where the signal is given by a global perturbation is, in this sense, protected against local perturbations, such as decoherence induced by the local environment. 

Since the order of the EP given by Eq.~(\ref{eq:ordermultipartite}) is less than the total algebraic multiplicity, there are additional eigenstates that must be entangled. This is highly remarkable, but it seems that it does not have any obvious physical consequences. It is, for instance, not related to the reported speeding up of entanglement in the proximity of an EP~\cite{LCA23}.

A simple equation [Eq.~(\ref{eq:rcageneral})] for the spectral response strength of the higher-order EP has been derived in terms of the response strengths of the leading-order EPs in the individual systems.
 
Ultimately, we unveiled the surprising phenomenon of EP-induced disentanglement, which occurs without any interactions between the individual systems.

We believe that our findings contribute significantly to the deeper understanding of non-Hermitian quantum systems and may help to advance practical applications in areas such as quantum optics, condensed matter physics, and open quantum systems.

\HL{The data that support the findings of this article are openly available~\cite{WCdata25}.}

\acknowledgments 
Valuable discussions with Z. Li, H. Schomerus, \HL{and E.~Winzer} are appreciated. 

%

\end{document}